\documentclass[]{tMOP2e}
\usepackage{pstricks}
\usepackage{psfrag}
\usepackage[T1]{fontenc}

\begin{document}

\doi{10.1080/0950034YYxxxxxxxx}
 \issn{1362-3044}
\issnp{0950-0340} \jvol{56} \jnum{18 \& 19} \jyear{2009} \jmonth{October}



\title{Factorization of numbers with truncated Gauss sums at rational arguments}

\author{S. W\"olk $^{\ast}$\thanks{$^\ast$Corresponding author. Email: sabine.woelk@uni-ulm.de
\vspace{6pt}}, C. Feiler and W. P. Schleich \\\vspace{6pt}  {\em{Institut f\"ur Quantenphysik, Universit\"at Ulm, Albert-Einstein Allee 11, D-89069 Ulm, Germany}}\\\vspace{6pt}\received{v2 transmitted June 2009} }

\maketitle

\begin{abstract}
Factorization of numbers with the help of Gauss sums relies on an intimate relationship between the maxima of these functions and the factors. Indeed, when we restrict ourselves to integer arguments of the Gauss sum we profit from a one-to-one relationship. As a result the identification of factors by the maxima is unique. However, for non-integer arguments such as rational numbers this powerful instrument to find factors breaks down. We develop new strategies for factoring numbers using Gauss sums at rational arguments. This approach may find application in a recent suggestion to factor numbers using an light interferometer [V. Tamma et al., J. Mod. Opt. in this volume] discussed in this issue.
\bigskip

\begin{keywords}Gauss sum; factorization; rational arguments
\end{keywords}\bigskip

\end{abstract}

\section{Introduction}
Newtonian mechanics originates only from those trajectories  in spacetime, which make the action an extremum. When we free ourselves  from this restriction  and  allow all trajectories, we arrive at quantum mechanics provided we introduce another selection criterion: although all trajectories contribute in a democratic way, that is with equal weight,  each trajectory carries a phase  determined by the action along this path. In this way we obtain the path-integral formulation \cite{wentzel:24,feynman:48} of quantum mechanics pioneered by Richard P. Feynman \footnote{It is interesting to note that already in 1924, that is before the advent of modern quantum mechanics introduced by Heisenberg and Schr\"odinger, Gregor Wentzel proposed the path integral in a paper entitled  "Zur Quantenoptik". Unfortunately, he never followed up on his idea.}. More freedom combined with a new rule in this enlarged realm  gives birth  to a new theory, which transcends  the original one.

On a much more elementary level, the factorization of numbers with the help of Gauss sums\cite{merkel:2009:a,merkel:2009:b} displays  a similar phenomenon. So far, we have restricted our analysis of  these sums to integer arguments. This requirement has provided us with a tool for the unique identification of the factors. In the present paper we allow more freedom and consider these sums at rational numbers. Needless to say, we immediately loose the possibility of identifying factors. However, when we introduce new rules for extracting factors the enlarged space of rational numbers  enables us  to factor numbers  more efficiently.

\subsection{Formulation of the problem}

Factorization of large integers is an important problem in network and security systems. The most promising development in this domain is the Shor algorithm\cite{shor:1994}, which uses the entanglement of quantum systems to obtain an exponential speed up.

Recently a different route towards factorization   was proposed. Based solely on interference,  an analogue computer  evaluates a Gauss sum. Several experimental implementations have been suggested \cite{dowling:1991, summhammer:1997,merkel:2009:a,merkel:2009:b}. In the meantime Gauss sum factorization has been demonstrated experimentally in various systems ranging from  NMR techniques \cite{mehring:2007,mahesh:2007,peng:2008}, cold atoms \cite{gilowski:2008,sadgrove:2008,sadgrove:2009}, ultra short laser pulses \cite{bigourd:2008,weber:2008} to classical light interferometry \cite{tamma:2009,tamma:2009:b}. The largest number factored so far had 17 digits. 

All experiments performed so far implement  the truncated Gauss sum 
\begin{equation}
{\cal A}_N^{(M)}(\ell)=\frac{1}{M+1}\sum\limits_{m=0}^M \exp\left(2\pi i\, m^2 \frac{N}{\ell}\right)
\label{eq:truncgauss}
\end{equation} 
where $N$ is the  number to be factored and the integer $\ell$ represents a trial factor.

The general idea of this approach is to find an observable of a physical system such as a spin, or the internal states of an atom, which is given by such a Gauss sum. Obviously a requirement for the successful application of Gauss sum factorization is that the parameters $\ell$, $M$ and $N$ can be chosen at will. In this way, we implement an analogue computer, that is the system calculates for us the 
truncated Gauss sum ${\cal A}_N^{(M)}(\ell)$. Unfortunately  the experiment does not provide us  directly with the factors but only with a "yes" or "no" answer to the question if $\ell$ is a factor. Indeed, the result of the experiment is:
\begin{center}
"$\ell$ is a factor of $N$" \qquad or \qquad
"$\ell$ is not a factor of $N$"\end{center}
As a consequence, we have to check every prime number up to $\sqrt{N}$. 
Most of the experiments performed so far in the realm of Gauss sum factorization had to check every trial factor $\ell$ individually. However, there already exists an experiment\cite{tamma:2009,tamma:2009:b} with classical light, where the Gauss sum ${\cal A}_N^{(M)}(\ell)$  is estimated simultaneously for every trial factor $\ell$. Here the number $\ell$ is encoded in the wavelength of the light and we deduce the  factor of $N$ from the spectrum of the light. In this way, we have the opportunity to obtain all factors simultaneously.

However, as often in life, every opportunity is accompanied by new challenges\footnote{One of the referees has speculated in his report that if each challenge would carry a phase we may find a new quantum mechanics of life. Indeed, Pascual Jordan\cite{jordan} and Erwin Schr\"odinger\cite{schroe:44} have argued that life might be intimately connected to quantum mechanics.}. Indeed, the spectrum in the experiment of Ref.\cite{tamma:2009,tamma:2009:b} is not represented  by the  Gauss sum at integer values of $\ell$ only. It is determined by the Gauss sum evaluated along the real axis.  So far, we have only used the Gauss sum of integers $\ell$ to factor $N$. As a consequence we throw away a lot of information contained in the full spectrum. The goal of the present  paper is to make the first step towards factorization with Gauss sums at real numbers: we discuss factorization with  truncated Gauss sum at rational arguments.

\subsection{Outline}

Our paper is organized as follows. In section \ref{sec:overview} we briefly summarize the theoretical and experimental work on Gauss sums. Then we analyse in section \ref{sec:congauss} the behavior of the truncated Gauss sum ${\cal A}^{(M)}_N$ for rational arguments. Our main interest is to obtain an answer to the following question:  is it possible to extract information about the factors of $N$ from peaks of the Gauss sum located at rational arguments. We show, that the answer is an emphatical "yes!" by providing  two methods. We conclude in section \ref{sec:summary} by a brief summary of our results.


\section{A family of Gauss sums\label{sec:overview}}
Gauss sums come in a large variety of forms and have already been analyzed  \cite{merkel:2009:a} in their possibilities to factor numbers. In this section, we give a brief introduction into the different types of Gauss sums and how to obtain factors from them. We conclude by summarizing the main theoretical and experimental results obtained so far.

\subsection{Continuous Gauss sum}
\begin{figure}[ht]
\begin{center}
\includegraphics[width=1\textwidth]{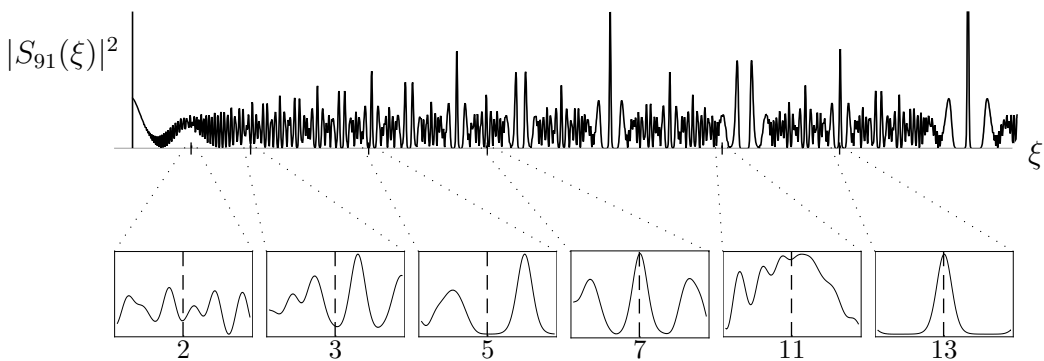}
\end{center}
\caption{
Factorization of $N=91=7 \cdot 13$ using the continuous Gauss sum $S_N(\xi)$ defined by Eq.(\ref{eq:s(xi)}).
On the top we present $|S_{91}(\xi)|^2$ in its dependence on $\xi$ for $1\leq\xi\leq 16$. 
On the bottom we magnify $|S_{91}(\xi)|^2$  in the vicinity of candidate primes. The pronounced maxima at the factors $7 \text{ and }13$ are clearly visible. In contrast, at non-factors the signal does not show any peculiarities. Here, we have used Gaussian weight factors $w_m$ centered at $m=0$ of width $\delta m=10$.
}
\label{fig:s(xi)}
\end{figure}

The Gauss sum 
\begin{equation}
S_N(\xi)\equiv
\sum\limits_{m= -\infty}^{\infty} w_m
\exp\left[2\pi i\left(m+\frac{m^2}{N}\right)\xi\right]\label{eq:s(xi)}
\end{equation}
with the weight factors $w_m$ depends on the continuous argument $\xi$ and contains an infinite number of terms. We obtain the factors of $N$ by testing if a maximum of this function is located at an integer value of $\xi$ as shown in Fig.\ref{fig:s(xi)}. The sum emerges\cite{mack} in the context of wave packet dynamics in the form of the autocorrelation function, or the excitation of a multi-level atom with chirped pulses \cite{merkel:2006}.

\subsection{Gauss sums with integer arguments}
The Gauss sum
\begin{equation}
S_N(\ell)\equiv
\sum\limits_{m=-\infty}^{\infty} w_m 
\exp\left[2\pi i\,m^2\frac{\ell}{N}\right]\label{eq:st_gauss}
\end{equation}
follows from the continuous Gauss sum Eq.(\ref{eq:s(xi)}), when we restrict the argument $\xi$ to integer values $\ell$. Since $\text{exp}\left[ 2\pi im\ell\right]=1 $ the term linear in $m$ disappears and only the fraction $\ell/N$ survives.

In Ref.\cite{merkel:2009:a} we transform $S_N(\ell)$ defined by Eq.(\ref{eq:st_gauss}) into the standard Gauss sum

\begin{equation}
G(\ell,N)\equiv\sum\limits_{m=0}^{N-1}\text{exp}\left[ 2\pi i\;m^2\frac{\ell}{N}\right].\label{eq:G(l,N)}
\end{equation}
In contrast to the Gauss sums $S_N(\xi)$ and $S_N(\ell)$ we now have a finite sum, which contains $N$ terms. This simplification emerges\cite{merkel:2009:a} from the fact, that the phase factors $\gamma_m\equiv\text{exp}\left[ 2\pi i m^2 \ell/N\right] $ are periodic in $m$ with period $N$, that is $\gamma_{m+N}=\gamma_m$.

For the factorization of numbers, it is useful to introduce the scaled square of the standard Gauss sum
\begin{equation}
g_N(\ell)\equiv \frac{1}{N}|G(\ell,N)|^2.
\end{equation}
The standard Gauss sum $G(\ell,N)$ can be calculated analytically \cite{prime numbers}.
Due to the division of $G(\ell,N)$ by $N$ we find the result 

\begin{equation}
g_N(\ell)= \text{gcd}(\ell,N)\label{eq:g_N}
\end{equation}
which gives us three possibilities to find a factor $q$ of $N$: 

(i) If $\ell= n \cdot q$, that is $\ell$ is an integer multiple of a factor, $g_N(n \cdot q)$ is larger than $1$, whereas for all other $\ell$  the function $g_N(\ell)$ is equal to unity.

(ii)The value of $g_N(\ell)$ is a factor of $N$, if $l$ is an integer multiple of this factor, that is $g_N(n \cdot q)=q$.

\begin{figure}
\begin{center}
\includegraphics[width=0.6\columnwidth]{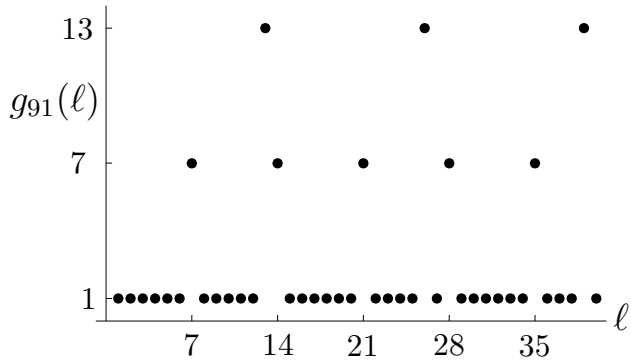}
\end{center}
\caption{
Factorization of $N=91= 7 \cdot 13$ using the scaled square of the standard Gauss sum  $g_{N}(\ell)\equiv|G(\ell,N)|^2/N$ defined by Eqs.(\ref{eq:G(l,N)}) and (\ref{eq:g_N}). In contrast to Fig.\ref{fig:s(xi)}, were we had the continuous variable $\xi$  we now restrict ourselves to  discrete arguments $\ell$. We recognize dominant maxima at integer multiples of the factors $7$ and $13$. Moreover,we  note the relation $g_N(n\cdot q)=q$ for integer $n$ and the factor $q$ of $N$. Indeed, the value of $g_N $ at an integer multiple of a factor is the factor.
}
\label{fig:st_gauss}
\end{figure}

(iii) The function $g_N(\ell)$ is periodic with the period $q$.\\
In Fig.\ref{fig:st_gauss} we illustrate this technique for $N=91=7 \cdot 13$. We note that the function $g_{91}(\ell)$ displays maxima at integer multiples of $7$ and $13$ with the values $7$ and $13$.

\subsection{Reciprocate Gauss sum}

The  Gauss sum 

\begin{equation}
{\cal A}^{(M)}_{N}(\ell)\equiv\frac{1}{M+1}\sum\limits_{m=0}^M 
\exp\left[2\pi i\, m^2\frac{N}{\ell}\right]\label{eq:trunc_Gauss}
\end{equation}
which is most popular in the context of experiments is again defined for  integer arguments but uses the fraction $N/\ell$, that is the reciprocate of the ratio appearing in $G(\ell,N)$. Morover, in ${\cal A}_N^{(M)}$ the summation extends only over $M$ terms, which is not necessarily identical to $N$. For this reason, this Gauss sum carries the name "Truncated Gauss sum".

\begin{figure}
\begin{center}
\includegraphics[width=0.65\columnwidth]{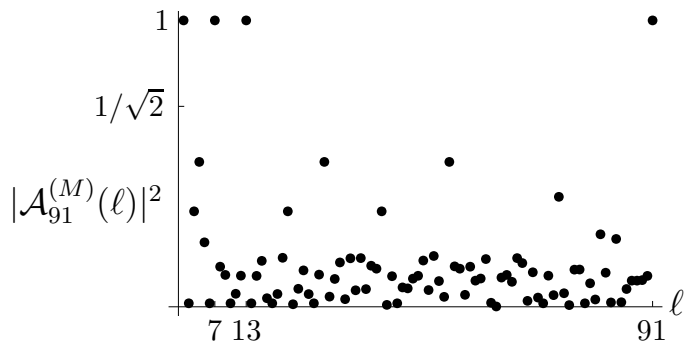}
\end{center}

\caption{
Factorization of $N=91=7 \cdot 13$  using the truncated Gauss sum ${\cal A}_N^{(M)}(\ell)$ defined by Eq.(\ref{eq:trunc_Gauss}). This function  assumes the value unity only for factors. For non-factors, the function falls below $1/\sqrt{2}$ provided $M$ is large enough. Here we have used $M=8$ for the evaluation of the Gauss sum.\label{fig:trunc_gauss}
}
\end{figure}

The truncated Gauss sum ${\cal A}^{(M)}_{N}(\ell)$ is of special interest, because it needs only a few summands to distinguish between factors and nonfactors. Provided the minimum number of terms  grows \cite{stefanak:2007,stefanak:2008:b} with $\sqrt[4]{N}$, the absolute value of ${\cal A}_N^{(M)}(\ell)$ is smaller than $1/\sqrt{2}$ for nonfactors and equal to unity for factors, as illustrated in Fig. \ref{fig:trunc_gauss}. For this reason all experiments on factorization with Gauss sums were performed with ${\cal A}_N^{(M)}$. 

\subsubsection{Experiments}
Many experiments are based on the interaction of laser pulses with  two-level systems. Here the fraction $N/l$ is encoded in the phases of the pulses. After the sequence of pulses the occupation probability  of  the excited level is given by the truncated Gauss sum. Such experiments were performed with cold atoms \cite{gilowski:2008} and liquids using Nuclear Magnetic Resonance \cite{mehring:2007,mahesh:2007,peng:2008}.

Other experiments relied on a Bose-Einstein-Condensate in an optical lattice \cite{sadgrove:2008,sadgrove:2009}. From the measured momentum distribution, the average energy, which is given by the truncated Gauss sum, can be obtained. This technique is very interesting, since it was claimed that the probability for higher momentum itself, can  be a more sensitive measurement.

All experiments so far share the problem \cite{jones}, that they have to calculate the ratio $N/l$ to determine the phases of the pulses. Fortunately there already exist some proposals for experiments with light, which do not suffer from this problem \cite{bigourd:2008, tamma:2009,tamma:2009:b}. The first proposal \cite{bigourd:2008} uses a sequence of interfering femto-second laser pulses with identical phases but variable separation between the pulses. 
The recent experiment \cite{tamma:2009,tamma:2009:b} employs an optical interferometer with variable optical path lengths. 
In both of the proposals, the truncated Gauss sum is calculated not only for integer values, but also for non-integers. This aspect has motivated us, to investigate the truncated Gauss sum at rational arguments as discussed in the next section.

\subsubsection{Methods to avoid ghost factors}
The truncated Gauss sums have the problem, that for some non-factors, their values are close to unity provided we do not take into account enough terms in the sum. We have called these factors ghost factors. 

There exist two techniques to improve factorization with  truncated Gauss sums and in particular, to decrease the number $M$ : (i) the first one relies on the exponential sums

\begin{equation}
{\cal A}^{(M,j)}_{N}(\ell)\equiv\frac{1}{M+1}\sum\limits_{m=0}^{M} 
\exp\left[2\pi i\, m^j\frac{N}{\ell}\right].
\end{equation}
Indeed, we avoid \cite{peng:2008,stefanak:2008:b} ghost factors, if $M$ is larger than, or of the order of $\sqrt[2j]{N}$  . 

(ii)The second method \cite{peng:2008, weber:2008} uses $M$ randomly chosen phases instead of   M consecutive phases. With this method the minimum number $M$ of terms necessary to factor a number grows only logarithmically. In this way it was possible to factor a 17-digits number  with only 10 pulses.


\section{More freedom:Truncated Gauss sum at rational  arguments\label{sec:congauss}}
So far, we have only considered the truncated Gauss sum ${\cal A}_N^{(M)}$  at integer arguments $\ell$. This restriction was motivated by the fact, that  ${\cal A}_N^{(M)}(\ell)$ is equal to unity, if and only if the integer trial factor $\ell$ is a factor of $N$. In the present section we show that  we can also gain information about the factors $p_k$ of $N$  from  Gauss sums at rational numbers.

\subsection{Three classes of rational arguments}
Such an extension of the theory has been made necessary by a recent experiment \cite{tamma:2009,tamma:2009:b} where all trial factors are checked simultaneously. This experiment involves the truncated Gauss sum not only for integer but also for rational numbers.

Indeed, the generalized truncated Gauss sum 
\begin{equation}
{\cal A}_N^{(M)}(\xi)\equiv\frac{1}{M+1}\sum\limits_{m=0}^M \exp\left(2\pi i\, m^2 \frac{N}{\xi}\right)
\label{eq:truncgausscon}
\end{equation}
for  the continuous variable  $\xi$ assumes the value 
${\cal A}_N^{(M)}(\xi)=1$
for rational numbers of the form $1/s,N/s$ and $p_k/s$ with s integer and $p_k$ factor of $N$.
The resulting  peaks have the information about the factors encoded either in their location, or in  their frequency of appearance, or do not contain any information at all. Obviously
the maxima at $\xi = p_k/s$ contain information since their locations are proportional to the factors $p_k$. Since the peaks at $\xi = 1/s$ are independent of $N$, they cannot give any hints about factors of $N$. Unfortunately, the analysis of the peaks at $\xi = N/s$ is not that straight forward. Nevertheless, we show in the next section, that it is possible to obtain information about the factors $p_k$.


\subsection{New selection criteria\label{sec:fast_analyse}}

The problem  with the truncated Gauss sum at rational arguments as a tool to factor numbers originates from the existence of additional peaks located at $\xi=1/s$ and $\xi=N/s$. On first sight, these maxima  do not seem to give us any information about the factors of $N$. However, in the present section, we outline two methods, that allow us to extract the factors from these additional peaks.

\subsubsection{Reduction and discretization of search interval}

If the Gauss sum would  not contain the peaks at $\xi=1/s$ or $\xi=N/s$, we could determine the factors immediately from the maxima at $p/s$. All we have to do in this case, is to search for a datapoint, where ${\cal A}_N^{(M)}(\xi)=1$. The value $\xi=p_k/s$ will then give us a factor $p_k$ of $N$. 

Hence, we have to develop a strategy how to eliminate unwanted peaks and/or identify those peaks which contain useful information.

From the peaks at $\xi = 1/s$ we cannot learn anything, since $1/s$ does not contain any information about the factors. Fortunately, it is easy to ignore those peaks, since  they are all located in the domain $\xi<1 $. For this reason, we restrict our domain of interest to arguments  $1<\xi $.

We now have to select from the forrest of peaks arising for $1<\xi $ the ones, which carry information about the factors. For this purpose, we discretize the curve obtained from the Gauss sum ${\cal A}_N^{(M)}$ with continuous argument $\xi$ by considering values of $\xi$, which are integer multiples of the minimal step size $1/s_0$.  Moreover,we confine the measurement range  to $1< \xi\leq \sqrt{N}/s_0$. This domain must contain peaks at  $\xi=p_k/s$, but there will be no peaks at $\xi=N/s$. The following consideration explains this observation.

We assume, that at a given multiple integer $\ell$ of the step size $1/s_0$,there exists a peak, which can be attributed to the ration $N/s$, where $N$ and $s$ do not share a commen factor. As a result, we have the identity 
\begin{equation}
\frac{N}{s}=\frac{\ell}{s_0}\quad\text{or}\quad \ell = \frac{s_0}{s}N.\label{eq:11}
\end{equation}
Since $s$ does not share a factor with $N$ and $\ell$ is an integer, $s$ has to be a divisor of $s_0$. However, this fact implies the inequality $s \leq s_0$ and from Eq.(\ref{eq:11}) we find $ N \leq \ell$. This inequality is in contradiction with our assumption of the domain $1< \xi\leq$ which provides us with the inequality $\xi =\ell/s_0 \leq \sqrt{N}/s_0$, that is $\ell \leq \sqrt{N}$.

\begin{figure}
\begin{center}
\includegraphics[width=0.8\textwidth]{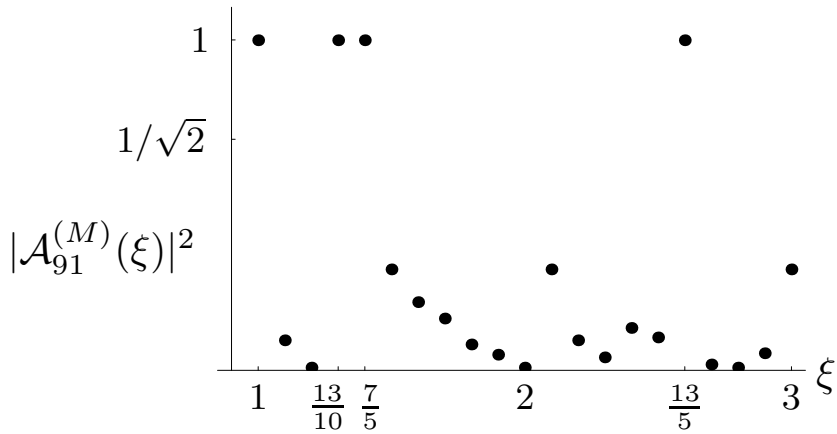}
\caption{Factorization using the truncated Gauss sum ${\cal A}_N^{(M)}(\xi)$ at rational arguments $\xi\equiv r/s$, with $r$ and $s$ integer, illustrated by the example $N= 91=7 \cdot 13 $.  Here we have  confined ourselves to  the range $1 \leq \xi \leq 3$ and have chosen a step size $1/s_0=1/10$. Despite of the fact that there are no factors in this domain, the Gauss sum assumes the value unity at the positions $\xi=13/10, 7/5$ and $13/5$, corresponding to the maxima at $\xi=p_k/s$ where $p_k$ is a factor of $N$. This analysis immediately suggests the factors $7$ and $13$.   Due to the reduced domain of interest and the discreteness of the observation points, we do not observe the peaks from arguments  $\xi=1/s$ or $\xi=N/s$ except the one located at $\xi=1$.}
\label{fig:gauss_kont}
\end{center}
\end{figure}

The reduction of the interval together with the discretization of the arguments of the truncated Gauss sum  ensure that all peaks appearing in the so-obtained diagram carry the information about the factors of $N$. Figure \ref{fig:gauss_kont} illustrates this method.

We conclude this discussion by recalling that in the standard approach of factoring numbers using Gauss sums at integer values, we have to search for factors at prime numbers smaller than $\sqrt{N}$. In the worst case, we have to test all prime numbers up to $\sqrt{N}$. In the present approach, we search for factors at integer multiples of $1/s_0$ in an interval up to $\sqrt{N}/s_0$. As a result, the present method scales in the same way as the standard one. However, the interval in which we perform the search is much smaller. It is in this sense that factorization with Gauss sums at rationa arguments might be more efficient than the corresponding one at integers. Moreover, it is appealing to use the full information contained in the Gauss sum.


\subsubsection{Degeneracy of ratios \label{sec:behavior_N_s}}
We now propose a second method to take advantage of the additional peaks in the Gauss sum at rational numbers. For this purpose  we analyse the series $N/s$ and $p_k/s$ and note, that   $p_k/s$ is contained in $N/s$. In other words, because $p_k$ is factor of $N$, some numbers $N/s$ can be expressed by $p_k/s'$. The number of different representations of the fraction $N/s$  depends of course on the factors of $N$. 

\begin{figure}
\begin{center}
\includegraphics[width=1\textwidth]{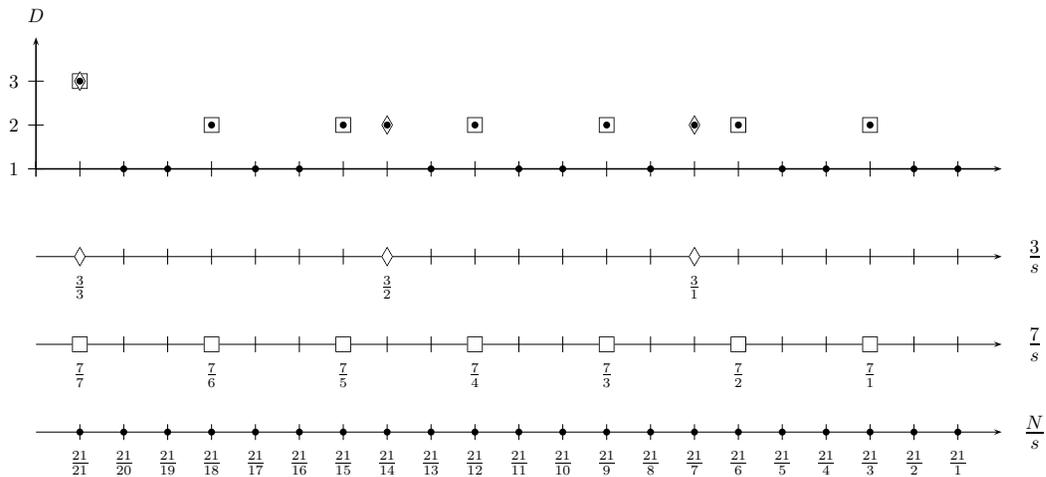}
\caption{Factorization of $N=21=3\cdot 7$ with the help of the number $D$ of degeneracies of the truncated Gauss sum ${\cal A}^{(M)}_N$ at rational arguments. 
For $N=p \cdot q$ this function assumes the value unity at  the rational numbers $p/s,q/s$ and $N/s$, marked on the corresponding axes by rhombs, squares and full dots, respectively. Fractions, which are connected with factors allow several representations. Indeed, the ratios $21/6$ and $7/2$  display the degree $D=2$ of degeneracy as indicated in the top by the full dot in the square. Only the ratio $21/21$ is identical to $3/3$ and $7/7$, giving rise to $D=3$. The degree of degeneracies displays a double periodicity determined by the two factors. For example, the squares with the dots have the period $3$, whereas the rhombs with the dots show the period $7$.}
\label{fig:haeufigkeit}
\end{center}
\end{figure}

In figure \ref{fig:haeufigkeit} we show  for the example of $N=3 \cdot 7=21$ the degree $D$ of degeneracy of these different representations of the same number, providing information about the factors. Indeed, the periods contained in  $D$ are the factors.

We conclude by briefly addressing the question of how to implement the degree of degeneracy of ratios in a quantum system. Many ideas offer themselves. In this context it suffices to name at least one. For this purpose we recall the phenomenon of quantum carpets \cite{berry}.
Here structures appear in the spacetime representation of a Schr\"odinger particle moving between two hard walls. Indeed, the design of a quantum carpet has its origin in the degeneracy of the so-called intermode traces \cite{kaplan}.
It is only a small additional step to conjecture that the degeneracy of the ratios manifests itself in different steepnesses of the canals. However, a more detailed analysis goes beyond the scope of the present paper and we refer to a future publication.

\section{Summary\label{sec:summary}}
The idea of factoring numbers using truncated Gauss sums relies on the fact, that these sums  assume the value unity if and only if the test factor is indeed a factor.
However, this one-to-one relation is only true when we restrict the arguments of the truncated Gauss sum to integer values. When we give up this restriction, we loose the possibility of uniquely identifying factors. In this case the truncated Gauss sum takes on the value of unity at many non-integer values, which obviously cannot be factors. However, this additional wealth of maxima also opens up new possibilities of factoring numbers. In the present paper we have introduced two methods, that allow us to obtain with the help of these on first sight useless peaks additional information about the factors. 

\section{Acknowledgment}
We thank M. Jakob, M. \v{S}tefa\'n\v{a}k and M. S.Zubairy for many fruitful discussions on this topic. This research was partially supported by the Max Planck Prize of WPS awarded by the Humboldt Foundation and the Max Planck Society.


\bibliographystyle{apsrev}


\end{document}